\documentclass[fleqn,usenatbib]{mnras}

\usepackage{newtxtext,newtxmath}

\usepackage[T1]{fontenc}

\DeclareRobustCommand{\VAN}[3]{#2}
\let\VANthebibliography\thebibliography
\def\thebibliography{\DeclareRobustCommand{\VAN}[3]{##3}\VANthebibliography}

\usepackage{graphicx}	
\usepackage{amsmath}	

\usepackage[pagewise,displaymath, mathlines]{lineno}

\newcommand{\divr}{\ensuremath{\nabla_r}}

\title[treatment of turbulent effects]{On the treatment of phenomenological turbulent effects in one dimensional simulations of core-collapse supernovae}

\author[Sasaki \& Takiwaki]{
Shunsuke Sasaki,$^{1,2}$\thanks{E-mail: sasaki.shunsuke.astro@gmail.com}
and Tomoya Takiwaki,$^{1,2}$
\\
$^{1}$Division of Science, National Astronomical Observatory of Japan, 2-21-1
Osawa, Mitaka, Tokyo 181-8588, Japan\\
$^{2}$School of Physical Sciences, Graduate University for Advanced Studies (SOKENDAI), 2-21-1 Osawa, Mitaka, Tokyo 181-8588, Japan\\
}

\date{Accepted 2023 December 4 . Received 2023 December 4 ; in original form 2023 July 6}

\pubyear{2023}

\begin{document}
\label{firstpage}
\pagerange{\pageref{firstpage}--\pageref{lastpage}}
\maketitle

\begin{abstract}
We have developed a phenomenological turbulent model with one-dimensional (1D) simulation based on Reynolds decomposition.
Using this method, we have systematically studied models with different effects of compression, mixing length parameters, and diffusion coefficient of internal energy, turbulence energy and electron fraction. With employed turbulent effects, supernova explosion can be achieved in 1D geometry, which can mimic the evolution of shock in the 3D simulations. We found that enhancement of turbulent energy by compression affects the early shock evolution. 
The diffusion coefficients of internal energy and  turbulent energy also affect the explodability. The smaller diffusion makes the shock revival faster. Our comparison between the two reveals that the diffusion coefficients of internal energy has a greater impact.
These simulations would help understand the role of turbulence in core-collapse supernovae.
\end{abstract}

\begin{keywords}
Core-collapse supernovae (304) -- Supernovae (1668); Hydrodynamical simulations (767) -- Supernova neutrinos (1666) -- Massive stars (732) -- Stellar convective zones (301)
\end{keywords}



\section{Introduction}\label{sec:introduction}
Massive stars undergo core-collapse supernova explosions in the final stage of their evolution. The energy source of the explosion is the enormous gravitational potential energy released from the implosion by their self-gravity.
Despite intensive and extensive studies in the past, the mechanism of the explosion has not yet been fully elucidated \citep[see][for the recent progress]{Kotake2012review, Burrows2013Review,Foglizzo2015review,Zhou2019,Burrows2020review,Mueller2020review}.
Complex phenomena such as neutrino radiative transport and convection play an essential role in the explosion, and
their numerical simulations require extremely sophisticated schemes.

Thanks to the enlarged computational resources, recently, self-consistent multi-dimensional simulations have been performed. Some of the models show the shock-revivals and explosions since neutrino-driven convection substantially enhances the neutrino heating \citep{Takiwaki2012,Hanke2013,Murphy2013,Abdikamalov2015,Lentz2015,Summa2016,Mueller2016,O'Connor2018_2D,Pan2018,Kuroda2018,Vartanyan2019,Nagakura2019,Melson2020,Burrows2020}.
Although those simulations are outstanding achievements of this field,
we have to point out that most of them have not been able to reproduce the typical explosion energy or the amount of synthesized $^{56} {\rm Ni}$ (\citealt{Murphy2019,Suwa2019}, however, see  \citealt{Bruenn2013,Nakamura2016,Bollig2021} for a few exceptions).
Still, we have not captured the nature of core-collapse supernovae, and we have missed something important in the simulations.

Encouraged by the partial success of the 3D simulations,
a variety of phenomenological 1D approaches have been used to mimic the 3D models and have linked the character of the progenitor and 
the explodability \citep{O'connor2011,Ugliano2012,Ertl2016,Sukhbold2016,Ebinger2020}.
The authors enlarge the neutrino luminosity and hence neutrino-heating to obtain the successful explosions since 
spherically symmetric 1D simulations cannot revive shocks \citep[e.g.,][]{Sumiyoshi2005}.
The studies found that several parameters govern the explodability: e.g., compactness, $\xi_{M}$; $M_4$ and $\mu_4$
\footnote{Here, 
$\xi_M \equiv \frac{M/M_\odot}{R(M) / 1000\,{\rm km}}$,
$M_4 \equiv \left.M\right|_{s=4}/M_\odot, \quad \mu_4 \equiv \left.\frac{{\rm d} m / M_\odot}{ {\rm d} r /1000\,{\rm km}}\right|_{s=4}$, where $M$, $R$ and $s$ are enclosed mass, radius and entropy per baryon, respectively.
$M_4$ and $\mu_4$ are evaluated at the radius where the entropy is $4\,{\rm k}_{\rm B}$.
}.
Those approaches are useful to compare the distribution of the observed properties \citep{Horiuchi2014,Pejcha2015} and to estimate 
diffusive supernova neutrino background integrating all neutrino flux from the past supernovae \citep{Lunardini2012,Horiuchi2018,Horiuchi2021,Kresse2021,Ashida2022,Ashida2023,Ekanger2022}. 

One current hot topic in this field is a discrepancy between the prediction in the phenomenological and self-consistent models. 
\cite{Burrows2020} have performed that 3D simulations, changing the progenitor mass systematically.
That and the following works \citep{Wang2022,Tsang2022} claim that the compactness does not govern explodability \citep[see also][for 2D studies]{Nakamura2015,Summa2016,O'Connor2018_2D}.
The previous phenomenological studies just enlarged the neutrino luminosities by hand.
Though this treatment would be motivated by the fact that turbulent effects make a longer residency timescale in the gain layer \citep{Murphy2008,Hanke2012,Nordhaus2010},
the convective effects of the self-consistent simulations would not be the same as the phenomenological one.

To solve the discrepancy between the self-consistent and phenomenological simulations, we need to theoretically understand the turbulent effect and make more sophisticated phenomenological models.
Here we introduce efforts to understand the turbulent convection in supernovae.
\cite{Murphy2011} have derived and summarized the equations for convective effects in this context.
\cite{Yamasaki_2006} have estimated how the shock revival becomes easier when employing strong energy transport due to convection.
\cite{Courch&Ott2015} and \cite{Radice2016} have concluded that the turbulent pressure pushes the shock outward and causes the explosion.
On the other hand, \cite{Mabanta2018} have claimed that the increased thermal pressure by the dissipation of turbulence makes a more substantial effect than the turbulent pressure. \cite{Kazeroni2018} have taken into account the size of the initial perturbations and the effect of the drag force. \cite{Mueller2019} has focused on the effect of turbulent viscosity.
We are under the middle way to the complete understanding of convective turbulence in core-collapse supernovae.
We should learn a lot from numerical studies in stellar evolution \citep[e.g.,][]{Arnett2015}.

Very recently, time-dependent simulations with phenomenological convective turbulence have been performed \citep{Mabanta2019,Couch2020,Mueller2019,Boccioli2021,Boccioli2022,Boccioli2023}.
Those try to model the convective turbulence in multi-dimensional simulations, and install the effects in one-dimensional simulations with several parameters, while the previous phenomenological studies just increased neutrino luminosity to trigger the explosion.
Following \cite{Mabanta2019}, we refer this method as \verb!1D+! \footnote{This approach is also called STIR, Supernova Turbulence In Reduced-dimensionality. This name is basically used in \cite{Couch2020} and \cite{Boccioli2021}.}.
Note that the \cite{Mabanta2019} try to establish global modeling of the convective turbulence where the turbulent properties are not determined by local hydrodynamic variables. On the other hand, other studies formulate it by local modeling \citep{Couch2020,Mueller2019,Boccioli2021}. In the following, we basically focus on the local modeling, however, we must stress the importance of the global modeling and further investigation on this topic.

These approaches not only would reproduce the dynamics close to the results of 3D simulations, but also are useful to determine the progenitor dependence of the explodability \citep{Couch2020,Boccioli2021,Zapartas2021}, taking advantage of its low computational cost as in other phenomenological approaches.
Predicting the neutrino luminosity, the frequency of the gravitational waves, and the optical light curves,
\verb!1D+! is also used in the context of multi-messenger observation of core-collapse supernova \citep{Warren2020,Barker2022}.

The purpose of this study is to understand the effects of turbulence in a current \verb|1D+| simulation. In particular, we perform a parameter survey to mimic the evolution of shock in a 3D simulation using the same progenitor model. We show that compression affects prompt convection in our 1D+ simulations. We also analyze the effect of turbulence parameters and show that the effect of diffusion is different from the effect proposed in previous studies. The weaker the diffusion, the more likely the explosion.
Our simulation method is introduced in Section~\ref{sec:numericalsetup} and Appendix~\ref{sec:formalism}.
In Section~\ref{sec:comp}, we show that the treatment of compression of turbulent energy has impacts on convection by comparing our results with a 3D simulation. 
In Section~\ref{sec:mixing}, we show the dependence of mixing length parameter on the shock evokution  comparing it to 3D simulation. In Section \ref{sec:diff}, we show the dependence of diffusion parameters on supernova explodability.

\section{Method}\label{sec:numericalsetup}

We employ the neutrino radiation hydrodynamic code of {\small 3DnSNe-IDSA} \citep{Takiwaki2016} and 
have newly implemented phenomenological convection effects to incorporate convection effects into 1D spherically symmetric simulations \citep[e.g.,][]{Mabanta2019,Mueller2019,Couch2020}. Based on Reynolds decomposition, the governing equations are written as follows:
\begin{linenomath*}
\begin{align}
    \partial_t \hat{\rho} +& \divr\left(\hat{\rho}\hat{v} + \langle \rho^{\prime} v^{\prime} \rangle  \right) = 0, \label{eq:massconsv}\\
    \partial_t \hat{\rho}\hat{v} +& \divr\left(\hat{\rho}\hat{v}\hat{v} + (\hat{P}+P_{\rm turb}) \right) \nonumber\\
    &= \frac{2\hat{P}+\frac{1-c}{c}P_{\rm turb}}{r}+\hat{\rho} g +S_\nu,\label{eq:momentum}\\
    \partial_t \hat{e}_{\rm tot} +& \divr\left[ (\hat{e}_{\rm tot}+\hat{P}+P_{\rm turb})\hat{v} +F_\epsilon+F_K\right]
    \nonumber \\
    &= \left( \hat{\rho} \hat{v} + \langle \rho^{\prime}
     v^{\prime} \rangle \right)  \mathrm{g} +Q_\nu,  \label{eq:totaleconsv}\\
    \partial_t e_{\rm turb} +& \divr \left[ e_{\rm turb}\hat{v} + F_K  \right] + 
    \dot{e}_{\rm HT}
    \nonumber \\
    & = 
    \langle \rho^{\prime} v^{\prime} \rangle g -\epsilon_{\rm dis}, \label{eq:turbulentecosv}\\
    \partial_t \hat{\rho}\hat{Y_e} +& \divr\left(\hat{\rho} \hat{Y_e}\hat{v} + F_{Y_e} \right) = \Gamma_\nu, \label{eq:efractioncosv}
\end{align}
\end{linenomath*}
where $\hat{\rho}$, $\hat{v}$, $\hat{e}_{\rm tot}$, $\hat{P}, \hat{Y}_e$, $\mathrm{g}$ are 
density, radial velocity, total energy, pressure, electron fraction, and gravitational acceleration, respectively.
We follow the notation of \cite{Mueller2019}, and $\hat{A}$ or $\langle A\rangle$ mean angle-average of the variable, $A$.
Prime is used for the deviation from the mean value.
The radial divergence, $\frac{1}{r^2}\frac{\partial}{\partial r} r^2$, is simply written as $\divr$.
Here $\hat{e}_{\rm tot}=\hat{\rho}\hat{\epsilon} + \hat{\rho}\hat{v}^2/2 + e_{\rm turb}$,
where $\hat{\epsilon}$ is specific internal energy and $e_{\rm turb}$ is turbulent energy.
In the first term of RHS in Eq.~\eqref{eq:momentum}, sum of $\theta\theta$ and $\phi\phi$ components of stress tensor appears \footnote{In the RHS of equation (26) of \cite{Couch2020}, they should have $2P/r$ that would be missing in the paper. On the other hand, Eq. (28) of \cite{Mueller2019} is correct. There $\partial_r P$ is used in LHS and is not $\frac{1}{r^2}\partial_r (r^2P_{rr}) $. Note that $\frac{1}{r^2}\partial_r (r^2P_{rr}) = \partial_r P_{rr} + \frac{2  P_{rr}}{r}$. In this case, $(P_{\theta\theta}+P_{\phi\phi}-2P_{rr})/r$ appears in RHS and is equal to zero in the isotropic case since $P_{rr} = P_{\theta\theta}=P_{\phi\phi}=P$. Finailly, no source term appears in RHS.} \citep[e.g. see Table~B.1 in][]{Gonzalez2007}. The term comes from the divergence of tensor in the spherical coordinate. Eq.~(130) in \cite{Stone1992a} denotes the derivation.

We explain the difference between the turbulence models of previous studies and ours, where
\begin{linenomath*}
\begin{align}
    e_{\rm turb}&=\frac{1}{2} {\rm tr}(\rho R),\label{eq:etubdef}\\
    P_{\rm turb}&=c\, {\rm tr}(\rho R).
\end{align}
\end{linenomath*}
Here $R$ is the turbulent components of the Reynolds tensor.  The constant $c$ determines how much fraction of the turbulent tensor contributes to the radial-radial component of turbulent pressure tensor. The relation between $e_{\rm turb}$ and $P_{\rm turb}$ is $P_{\rm turb} = 2 c e_{\rm turb}$. 
The pressure due to turbulence is assumed as $P_{\rm turb} = e_{\rm turb}$ with reference to \cite{Couch2020} and \cite{Mueller2019}. \cite{Couch2020} adopted the same closure  of \cite{Mabanta2019} as Reynolds tensor, which by our definition is $R_{rr} = 2 R_{\theta \theta} = 2 R_{\phi \phi} $.  We assume those for turbulent pressure and set $c=0.5$ in order to be consistent with notation of \cite{Couch2020} and \cite{Mueller2019}. Similar to the pressure tensor, we also have turbulent pressure in RHS of Eq.~\eqref{eq:momentum}, $\frac{\rho R_{\theta\theta}+\rho R_{\phi\phi}}{r}=\frac{(1-c){\rm tr}{(\rho R)}}{r}=
\frac{1-c}{c}\frac{P_{\rm turb}}{r}$. This term is ignored in the previous studies
\footnote{In \cite{Couch2020}, $\rho(R_{\theta\theta}+R_{\phi\phi})/r=P_{\rm turb}/r$ is ignored in Eq.~(26). In \cite{Mueller2019}, $\rho(R_{\theta\theta}+R_{\phi\phi}-2R_{rr})/r=-P_{\rm turb}/r$ should appear in Eq.~(28).
}.

The effects of neutrino interaction are written as $Q_\nu$, $\Gamma_\nu$, and $S_\nu$.
We calculate $Q_\nu$, $\Gamma_\nu$, using Isotropic Diffusion Source Approximation \citep[IDSA, ][]{Liebendorfer2009,Takiwaki2014,Kotake2018}. $S_\nu$ is ignored in the approximation since the mass of nucleons are high compared to the energy of the neutrinos, and the momentum of the matter is hardly changed by the neutrino scattering. On the other hand, the effect of the collision is included in the neutrino transport.

This paper argues the impact of $\dot{e}_{\rm HT}$ in Eq.\eqref{eq:turbulentecosv} (HT is abbreviation of Hydrodynamic enhancement of the Turbulent energy).
\cite{Mueller2019} and \cite{Couch2020} use $\dot{e}_{\rm HT} = P_{\rm turb} \frac{1}{r^2}\partial_r r^2\hat{v}$ and $\dot{e}_{\rm HT} = P_{\rm turb} \partial_r \hat{v}$, respectively.
Though those looks similar, the meanings are slightly different.
The former indicates the enhancement of the turbulent energy is due to the compression and the latter indicates that the shear is important.
The origin of this term is actually the shear (see Eq.~(7) of \cite{Murphy2011}, the term is called shear production) \footnote{The original $\dot{e}_{\rm HT}$ is $\sum_{i,j} \langle \rho v_i^\prime v_j^\prime\rangle \partial_j \langle v_i \rangle$. Following Eq.~(23b) of \cite{Kuhfuss1986}, $\langle \rho v_i^\prime v_j^\prime\rangle \sim -\rho \mu\left[ \partial_i v_j^\prime + \partial_j v_i^\prime-\frac{2}{3}\nabla\cdot v^\prime\right]+\frac{\delta_{ij}}{3}\rho |v^\prime|^2$. Taking $\mu=0$, only the diagonal terms remain and $\dot{e}_{\rm HT}=\sum_{i} \frac{\delta_{ii}}{3}\rho |v^\prime|^2 \partial_i \langle v_i \rangle = \frac{\rho |v^\prime|^2}{3} \nabla \cdot  \langle v \rangle = P_{\rm turb}\nabla \cdot \langle v \rangle$ where $\frac{\rho |v^\prime|^2}{3}= P_{\rm turb}$ for isotropic turbulence. While  $\dot{e}_{\rm HT}$ is not simply proportional to $\nabla \cdot \langle v \rangle$ in the case of anisotropic turbulence, we assume $\dot{e}_{\rm HT}=P_{\rm turb}\nabla \cdot \langle v \rangle$ for simplicity. }.
However, the total energy conservation is not achieved by the latter expression and the former expression ensures the energy conservation (see Appendix~\ref{sec:formalism}).
Suming up $v\cdot \nabla P_{\rm turb}$ and $P_{\rm turb} \nabla \cdot v$ in Eq.~\eqref{eq:rynld_ekin} and \eqref{eq:rynld_etub}, we obtain $\nabla \cdot( P_{\rm turb}v)$ in Eq.~\eqref{eq:totaleconsv}. If $\dot{e}_{\rm HT} \neq P_{\rm turb} \frac{1}{r^2}\partial_r r^2\hat{v}$, consequently Eq.~\eqref{eq:totaleconsv} should be acoordingly modified. In this paper, we tolerate this inconsistency.
We need to compare these pros and cons employing this term. See Section~\ref{sec:comp} for the detail.

We estimate multi-dimensional turbulent effect using mixing length theory as follows.
The effect is introduced from  10\,ms after core bounce. 
At first, the mixing length $\Lambda$ is
\begin{linenomath*}
\begin{equation}
\Lambda = \alpha_\Lambda H_P = \alpha_\Lambda \frac{\hat{P}}{\hat{\rho} g},
\end{equation}
\end{linenomath*}
where $H_P$ is scale height of pressure and $\alpha_\Lambda$ is dimensionless parameter, which is also used in \cite{Couch2020} and \cite{Mueller2019}.

In the governing equations, 
$F_{u},u = \epsilon ,Y_e , K$ are the turbulent flux of internal energy, turbulent energy and electron fraction, $Y_e$, respectively.
To evaluate them, we take the gradient diffusion approximation. 
In the mixing length theory, the diffusion coefficients are defined by
the mixing length, $\Lambda$ and the turbulent velocity, $v_{\rm turb}$:
\begin{linenomath*}
\begin{equation}
D_u = \alpha_u v_{\rm turb} \Lambda \quad (u = \epsilon ,Y_e , K) \label{eq:diff},
\end{equation}
\end{linenomath*}
where $\alpha_u$ is the parameter for each of the three diffusion coefficients. Approximately, the turbulent velocity is evaluated as 
\begin{linenomath*}
\begin{equation}
    v_{\rm turb}= \sqrt{e_{\rm turb}/\hat{\rho}}.\label{eq:vturbdef}
\end{equation}
\end{linenomath*}
The three fluxes due to turbulence can be written in terms of their respective diffusion coefficients as follows
\begin{linenomath*}
\begin{align}
    F_\epsilon =& -\hat{\rho} D_{\epsilon} \left( \frac{\partial \hat{\epsilon}}{\partial r} + \hat{P} \frac{\partial}{\partial r} \left( \frac{1}{\hat{\rho}} \right) \right),\label{eq:dif_flux_e}\\
    F_K =& -\hat{\rho} D_K \partial_r v^2_{\rm turb},\label{eq:dif_flux_k}\\
    F_{Y_e} =& -\hat{\rho} D_{Y_e} \partial_r Y_e.
\end{align}
\end{linenomath*}
Those terms are consistent with those appearing in Eqs. (29)--(31) of \cite{Mueller2019}.

The source term of the turbulent energy in Eq.~\eqref{eq:turbulentecosv} is expressed as
\begin{linenomath*}
\begin{equation}
\langle \rho^{\prime} v^{\prime} \rangle\cdot g = \hat{\rho} v_{\rm turb} \omega^2_{\rm BV} \Lambda, \label{eq:sourse_turb}
\end{equation}
\end{linenomath*}
where $\omega_{\rm BV}$ is the Brunt–Väisälä frequency, which is defined as:
\begin{linenomath*}
\begin{align}
    \omega_{\rm BV}^2 &= \frac{C_{\rm L}}{\rho} |\mathrm{g}_{\rm eff} |,\label{eq:omega_bv}\\
    C_{\rm L} &=  \frac{\partial \rho}{\partial r} - \frac{1}{c_{\rm s}^2} \frac{\partial P}{\partial r}  \nonumber\\
              &= \left( \frac{\partial \rho}{ \partial s} \right)_{Y_e , P} \frac{\partial s}{ \partial r} +\left( \frac{\partial \rho}{ \partial Y_e } \right)_{s,P} \frac{\partial Y_e}{\partial r} \label{eq:entropy-gradient},
\end{align}
\end{linenomath*}
where $C_{\rm L}$ represents the density difference from the surroundings when the fluid element is adiabatically displaced by ${\rm d} r$. 
$\mathrm{g}_{\rm eff}$ is effective gravitational acceleration,
\begin{linenomath*}
\begin{equation}
    g_{\rm eff} = g + \hat{v}\frac{\partial \hat{v}}{\partial r} .
\end{equation}
\end{linenomath*}

This term represents the apparent gravitational acceleration in a fluid stationary system mentioned in Eq.(16) of~\citep{Couch2020}.
In our notation, $\omega_{\rm BV}^2$ takes a positive value 
in the convectively unstable region where the entropy, $s$, or $Y_e$ gradients are negative \footnote{Depending on equation of state, $\left(\frac{\partial \rho}{ \partial Y_e } \right)_{s,P}$ in Eq.~\eqref{eq:entropy-gradient} could be positive and in that case, the negative $Y_e$-gradient does not imply convectively unstable \citep[e.g.,][]{Bruenn2004}.}.
Since small turbulent seed is required to turbulent energy source, we assume that the minimum value of turbulent energy source following Eq.~(23) in \cite{Couch2020} is 
\begin{equation}
    (\langle \rho^{\prime} v^{\prime} \rangle\cdot g )_{\rm min} = \hat{\rho} \omega^4_{\rm BV} \Lambda^2 {\rm d} t .
\end{equation}
Note that \cite{Mueller2019} use a different handling, see their Eq.~(47).

In this study, we focus on the role of turbulence in the gain region and 
switch off this term at $\hat{\rho}>10^{11}\,{\rm g/cm^3}$ not to include the convection in PNS. In order to suppress turbulence generation at the shock discontinuity, $\omega_{\rm BW}$ is set to zero near the shock by hand.
Note that this term naturally appears in Eq. \eqref{eq:totaleconsv}.
Since we ignore $\langle \rho^{\prime} v^{\prime} \rangle$ in Eq.~\eqref{eq:massconsv}, the treatment is not consistent and is criticised by \cite{Mueller2019}. The total energy including gravitational potential energy, does not conserve in this formalism.
However, as discussed in Appendix \ref{sec:formalism}, the treatment of \cite{Mueller2019} invokes another problem.
In addition, Mabanta and Murphy have stated in private communication that the addition of this term will not have a significant impact.
We prefer to stick the current formalism.

Eq.~\eqref{eq:sourse_turb} is a major assumption of our study.
\cite{Murphy2013} claim that the generation term is proportional to the neutrino luminosity, not the negative entropy gradient itself, which is significantly altered by turbulent mixing. Since our simulations solve time-dependent fluid dynamics with neutrino radiative transport, we adopt Eq.~\eqref{eq:sourse_turb} for simplicity as a first step. This treatment should be updated in the future. We discuss this issue again in Section~\ref{sec:diff}.

We evaluate the turbulent dissipation, $\epsilon_{\rm dis}$, following Kolmogorov’s hypotheses \citep[see Section~4.1 in][]{Murphy2011}. The turbulent energy is injected at the largest scales and cascades to smaller scales. Consequently, the largest scale of the eddy, $\Lambda_{\rm diss}$, governs the rate of the dissipation:
\begin{linenomath*}
\begin{equation}
    \epsilon_{\rm dis} = \hat{\rho} \frac{v^3_{\rm turb}}{\Lambda_{\rm diss}}.
\end{equation}
\end{linenomath*}
Following \cite{Mueller2019}, we incorporates the effect of the overshooting:
\begin{linenomath*}
\begin{equation}
\Lambda_{\rm diss} = \max\left[ 
\min\left(
\Lambda,\sqrt{\frac{v_{\rm turb}^2}{\max\left[-\omega_{\rm BV}^2,0\right]}}
\right)
,\delta \right],
\end{equation}
\end{linenomath*}
where $\delta =10^{-10}{\,}{\rm cm}$.

The details of numerical treatment on the governing equations are written in \cite{O'Connor2018}.
The numerical flux is calculated by a HLLC solver \citep{Toro1994}.
A piecewise linear method with the geometrical correction of \cite{Mignone2014} is used to reconstruct variables at the cell edge, where a modified van Leer limiter is employed to satisfy the condition of total variation diminishing (TVD).
The computational grid is comprised of $512$ logarithmically spaced, radial zones that cover from the center up to the outer boundary of $5\times10^8$\,cm. The effect of the GR potential is included by Case A of \cite{Marek2006}.

The setup for the microphysics is the same to {\it set-all} of \cite{Kotake2018}, which employ the latest neutrino reaction rate, e.g., the medium effect is included \citep{Horowitz2017}. We use the EOS of \cite{Lattimer1991}
and set the incompressibility
parameter $K=220\,{\rm MeV}$, which can sustain $2.05\,M_\odot$ NS mass \citep{Fischer2014} that is comparable to 
the observed maximum masses of neutron stars ($1.97\,M_\odot$ in \citealt{Demorest2010}, $2.01\,M_\odot$ in \cite{Antoniadis2013},  $2.08\,M_\odot$ in \citealt{Cromartie2020,Fonseca2021}).

\section{Results}\label{sec:results}
This section presents the outcomes of hydrodynamic simulations, which utilize two turbulence models. The first model, the non-compressive case, ignores the enhancement of turbulent energy due to compression, while the second model, the compressive case, takes this enhancement into account. We systematically explore the impact of changes in the mixing length parameter, $\alpha_\Lambda$, and the three diffusion coefficients, $\alpha_\epsilon$, $\alpha_K$, and $\alpha_{Y_e}$. For all models, $12M_\odot$ progenitor model of \cite{WooslyHegar2007} is used as the initial condition.

The initial part of the section examines the effect of compression on turbulence, displaying the contrasts in the shock evolution between the non-compressive and compressive cases. We also present the time-dependent variation of the source term of turbulence energy and the distribution of turbulent velocity to illustrate the influence of compression in our turbulence models.

The subsequent section employs the non-compressive models and showcases the explosion dynamics while altering the mixing length parameter. We exhibit the differences in shock evolution for varying mixing length parameters and present the variations in the velocity distribution of turbulence to emphasize the impact of this parameter.

Finally, the last section illustrates how the dynamics change as a result of diffusion parameters. Analogous to the previous section, we present the differences in shock evolution and velocity distribution of turbulence to demonstrate the effects of diffusion parameters.

\subsection{Compressive enhancement of turbulence }\label{sec:comp}
In Section~\ref{sec:numericalsetup}, we introduced several types of terms that appear in the turbulent energy equation, Eq.\eqref{eq:turbulentecosv}. Here we just write it again for the readers' convenience.
\begin{linenomath*}
\begin{align}
      \partial_t e_{\rm turb} +& \divr \left[ e_{\rm turb}\hat{v} + F_K  \right]  
     = -\dot{e}_{\rm HT}+
    S_{\rm turb} -\epsilon_{\rm dis}\label{eq:turbulentecosv-2}
\end{align}
\end{linenomath*}
We compare two turbulent models, non-compressive case and compressive case.
For the non-compressive case, we assume $\dot{e}_{\rm HT}=0$, while for the compressive case,
we use $\dot{e}_{\rm HT} = P_{\rm turb} \frac{1}{r^2}\partial_r r^2\hat{v}$.
We implement this term as follows,
\begin{linenomath*}
\begin{equation}
\begin{split}
\dot{e}_{\rm HT,i} =& P_{{\rm turb},i}\frac{3}{
r_{i+1/2}^3-r_{i-1/2}^3}\\
&\times \left(
r^2_{i+1/2}\hat{v}_{i+1/2,L} -r^2_{i-1/2}\hat{v}_{i-1/2,R} 
\right),
\end{split}\label{eq:EHTimplementation}
\end{equation}\end{linenomath*}
where the subscript $i$ means $i$-th grid.
The radial velocity, $\hat{v}$, should be evaluated in the cell edge (represented by $i+1/2$ or $i-1/2$).
Since we use 2nd-order interpolation to obtain the numerical flux, we utilize the interpolated value.
The subscript $R$ and $L$ denotes the direction of the interpolation. 


\begin{figure}
    \centering
    \includegraphics[width=.99\linewidth]{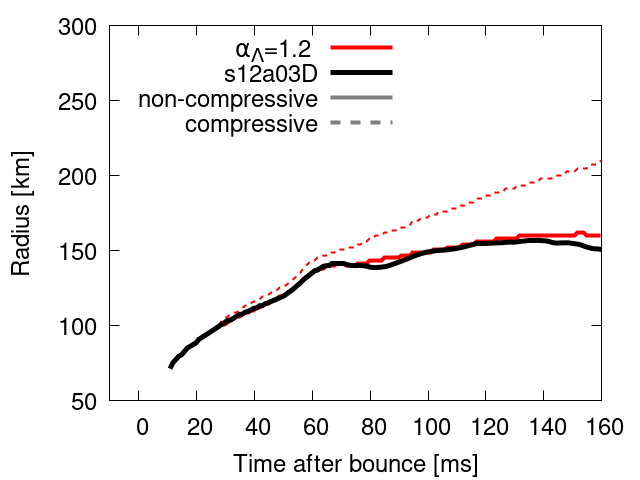}
    \caption{Comparison of the shock evolution in the compressive and the non-compressive cases. The vertical axis is the shock radius, and the horizontal axis represents the time after the bounce. The red solid curve corresponds to the non-compressive case, and the red dashed curve does the compressive case.  The black line denotes the time evolution of the average shock radius in the 3D simulation using the same progenitor.}
    \label{fig:CompShock}
\end{figure}

 Figure~\ref{fig:CompShock} displays the evolution of the shock for the model without compression (non-compressive case) and with compression (compressive case) . The vertical axis is the shock radius and the horizontal axis represents the time after core bounce. The solid line corresponds to the non-compressive case and the dashed line does compressive case.  The black line denotes the time evolution of the average shock radius for the 3D simulation using the same progenitor. The angular grid number of the 3D model is $64 \times 128$ \citep[see][for the numerical setup. The emplyoed progenitor is different from this paper]{Takiwaki2016}. In the 3D model, the turbulence naturally occurs, and the phenomenological effect is not included.
 We define the shock radius based on Eq.~(72) of \cite{Marti&Mueller1996}, i.e., a discontinuous surface where the pressure difference is greater than the value of the pressure in the neighborhood, and the velocity is converging. The value of entropy is also used to make it more smooth in time.

Based on the results, we can conclude that the non-compressive case (red solid) is closer to the evolution of the 3D model (black bold) than the compressive case (red dashed). 
Compressive case shows different behavior from the 3D model from 50\,ms, and radius of the shock is significantly deviates from the 3D model as time passes; for instance, 50\,km of the difference appears in 160\,ms.
The turbulent energy is considerably increased by the compression term. 
To reproduce the early shock trajectory, accurately, it is preferable to ignore the effect of $\dot{e}_{\rm HT}$.
One may expect that the compressive case with smaller $\alpha_\Lambda$ can fit to the 3D model. If we use the parameter that makes the shock radius at 100\,ms consistent with 3D, we cannot find shock revival and the later evolution is not consistent with 3D (see Figure~\ref{fig:shock-mix} for the evolution in the later epoch).

\begin{figure}
    \centering
    \includegraphics[width=.99\linewidth]{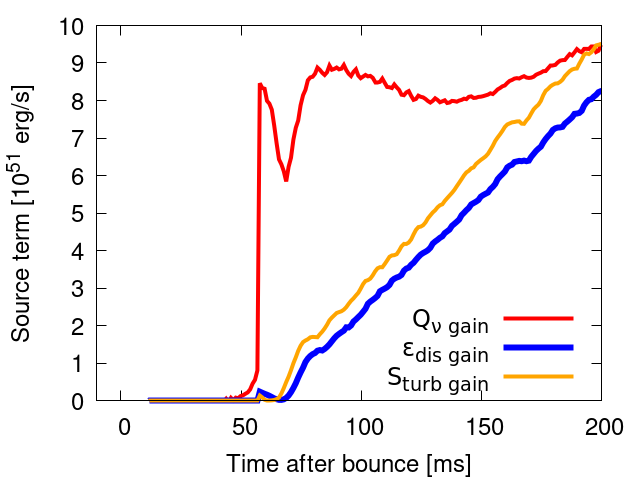}
    \includegraphics[width=.99\linewidth]{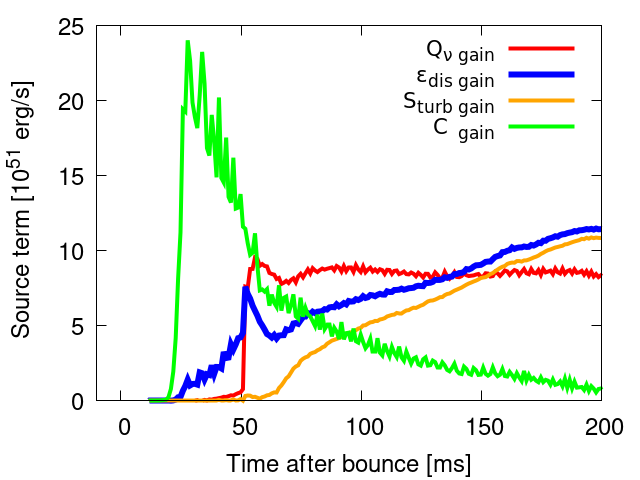}
    \caption{Time evolution of source terms of turbulent energy with the non-compressive and compressive cases. The source terms in the turbulent energy equation is integrate over the gain region, see Eqs.~\eqref{eq:S-gain}--\eqref{eq:Qnu-gain}.The terms are shown as a function of the postbounce time in ${\rm ms}$. The top panel and bottom panels represent the non-compressive case and the compressive case, respectively. Orange, blue, and green indicate turbulent energy source rate, turbulent energy dissipation rate, and turbulent energy increase rate due to turbulence compression in the gain region, respectively. Red is the neutrino heating rate. }
    \label{fig:CompGains}
\end{figure}

To understand the detail of the two models, we compare the generation and dissipation term of the turbulent energy.
Figure \ref{fig:CompGains} shows neutrino heating rates and the source terms in the governing equation for turbulent energy in the gain regime. The vertical axis is source terms in the unit of $10^{51}\,{\rm erg/s}$ and the horizontal axis is postbounce time in ${\rm ms}$. The top panel and bottom panels represent non-compressive case, and compressive case, respectively. Orange, blue, and green indicate the turbulent energy source rate, turbulent energy dissipation rate, and turbulent energy increase rate due to turbulence compression in the gain region, respectively. The definition of each term is as follows
\begin{linenomath*}
\begin{align}
    S_{{\rm turb \, gain}} &= 4 \pi \int_{r_{\rm gain}}^{r_{\rm shock}} \left(\rho v_{\rm turb} \omega^2_{\rm BV} \Lambda \right) r^2 {\rm d} r,\label{eq:S-gain}\\
    \epsilon_{{\rm dis \, gain}} &= 4 \pi \int_{r_{\rm gain}}^{r_{\rm shock}} \rho \frac{v^3_{\rm turb}}{\Lambda_{\rm diss}} r^2 {\rm d}r ,\label{eq:Diss-gain}\\
    C_{{\rm gain}} &= 4 \pi \int_{r_{\rm gain}}^{r_{\rm shock}} \left(- P_{\rm turb} \frac{1}{r^2}\partial_r\left( r^2 \hat{v}\right)\right) r^2 {\rm d} r \label{eq:C-gain}.
\end{align}
\end{linenomath*}
In the figure the red curve is the integrated neutrino heating rate, i.e.,
\begin{linenomath*}
\begin{align}
    Q_{\nu \, {\rm gain}} &= 4 \pi \int_{r_{\rm gain}}^{r_{\rm shock}} Q_\nu r^2 {\rm d} r.\label{eq:Qnu-gain} 
\end{align}
\end{linenomath*}

In the governing equations, $S_{\rm turb \, gain}$ increase the turbulent energy and $\epsilon_{\rm dis \, gain}$ decrease the energy. We expect that $S_{\rm turb \, gain}$ is larger than the $\epsilon_{\rm dis \, gain}$ before the system becomes equilibrium.
In the top panel, we confirm $S_{\rm turb \, gain} > \epsilon_{\rm dis \, gain}$.
\cite{Murphy2013} reported that the turbulence in the gain region is maintained by the neutrino heating and 
$Q_\nu > S_{\rm turb \, gain}$ is also expected.
In the top panel, $Q_\nu > S_{\rm turb \, gain}$ is found before 170\,ms and those terms remains comparable after that time. These features are roughly consistent with \cite{Murphy2013}.
Here we add a caveat.
In our model, the source term, $S_{\rm turb \, gain}$ indirectly related to the neutrino heating. The neutrino heating is stronger at the bottom of the gain region and is weaker near the shock. This makes negative entropy gradient and 
that determines Brunt–Väisälä frequency,
see Eqs.~\eqref{eq:omega_bv}--\eqref{eq:entropy-gradient}.
Essentially, the neutrino heating accounts for the generation of the turbulent energy.
Our treatment is valid in the liner phase of turbulent growth.
In the non-linear phase, the entropy gradients is flatten by the turbulence and the our treatment may underestimate the generation term. To avoid the problem, the source term in \cite{Mabanta2019} is directory related to the neutrino heating. We may try to implement this method in the future.

In contrast to the non-compressive case, the compressive case poses several issues.
In the bottom panel of Figure \ref{fig:CompGains}, the contribution of source terms in the gain region are shown in the compressive case.
First, it is unlikely that
the dissipation term, $\epsilon_{\rm dis \, gain}$, is larger than the source term, $S_{\rm turb \, gain}$ in all time, $0$--$200$\,ms. 
Instead of $S_{\rm turb \, gain}$, the compression term $C_{{\rm gain}}$ plays dominant production role in the initial phase, 0--50\,ms.
Here the turbulent energy of the seed turbulence is amplified by compression near the shock. 
In this phase, the energy gain is much larger than the neutrino heating.
It is unexpected and contrary to the situation observed in multi-dimensional simulations.
After 50\,ms, $C_{{\rm gain}}$ becomes smaller but still it is larger than $S_{\rm turb \, gain}$ until 100\,ms. Therefore the turbulent energy generation is overestimated in the compressive case until 100\,ms.
The generated turbulent enegy and turbulent pressure strongly pushes the shock and the shock trajectory in the compressive case is overestimated compared to the 3D model (see Figure~\ref{fig:CompShock}).
Our governing equations for the compressive case is similar to a model in \cite{Mueller2019}, the model is called non-conservative, no turbulent viscosity.
Surprisingly, their calculation did not exhibit an overestimation of the shock position, and we are uncertain about the underlying reasons for this.
It could potentially be due to the implementation of the term, see Eq.~\eqref{eq:EHTimplementation} for our particular case.

\begin{figure}
    \centering
    \includegraphics[width=.99\linewidth]{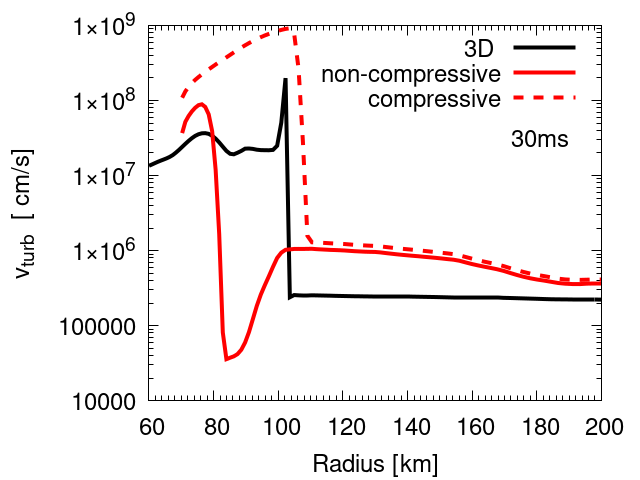}
    \caption{Radial profile of turbulent velocity at 30\,ms postbouce. The horizontal axis is the radius, and the vertical axis is the turbulent velocity. The solid red line corresponds to the non-compressive case, and the dashed red line is the compressive case. The black line is the result of the same time in the 3D simulation.
    }
    \label{fig:CompVturb}
\end{figure}

To understand that turbulent energy enhancement is different in the compressive and non-compressive cases, we
conducted a detailed investigation into the radial profile of turbulence in the early phase.
Figure~\ref{fig:CompVturb} shows turbulent velocity distribution at 30\,ms postbouce. The horizontal axis is the radius, and the vertical axis is the turbulent velocity. The solid red line corresponds to the non-compressive case, and the dashed red line is the compressive case. The black line is the result for the same time in 3D, which
is derived by Reynolds decomposition, which is a way to decompose physical quantities into average and perturbations,
$A = \langle A\rangle  + A^\prime, $
where $\langle A \rangle$ is the solid angular average and is defined as
$\langle A\rangle = \frac{1}{4 \pi} \iint A(r,\theta,\phi) \sin{\theta} {\rm d}\theta {\rm d}\phi$.
The velocity of the 3D turbulence shown in the figure is defined as
\begin{linenomath*}
\begin{equation}
    v_{\rm turb \, 3D} = \sqrt{\frac{R_{rr} + R_{\theta \theta} +R_{\phi \phi}}{2} },
\end{equation}
\end{linenomath*}
where
$R_{ij} = (v_i - \langle v_i \rangle)(v_j - \langle v_j\rangle)$ is Reynolds tensor. The factor 2 in RHS comes from Eqs.~\eqref{eq:etubdef} and \eqref{eq:vturbdef}, i.e.,
$v_{\rm turb}=\sqrt{{\rm tr}(R)/2}$.

The turbulent velocity of the compressive case (dashed red) is clearly higher than that of 3D because the turbulent energy is amplified near the shock due to compression. 
The seed turbulence is $\sim 10^{5-6}$\,cm/s outside the shock and undergoes enhancement due to the shock-compression in the early phase, as described in Figure~\ref{fig:CompGains}.
Comparatively, the turbulent pressure is excessively amplified in the compressive models, thereby driving the shock forward. 
The resultant radial profile diverges significantly from the 3D models.
To mitigate this issue, we can omit the compression term. In the non-compressive case the source terms are reasonable (see Figure~\ref{fig:CompGains}), and the shock evolution is more consistent with 3D (see Figure~\ref{fig:CompShock}).
As an inherent trade-off of this artificial hudling, the turbulent velocity near the shock is underestimated.
In Figure~\ref{fig:CompVturb}, the velocity at 80--100\,km in the non-compressive case (solid red) is smaller than the 3D case.
On the evolution of the shock, this underestimate of the turbulent velocity is less harmful than the overestimate in the compressive case.
In the future, we should investigate the method that can mimic 3D profile.
The inconsistency between 1D+ model and 3D model starts from the seed turbulent velocity at the pre-shocked region. 
As discussed in \cite{Kazeroni2018}, the growth rate of the convective velocity also depends on the radial velocity and wavenumber of the perturbation. The rate has the maximum in some wavenumber with zero velocity, that is Brunt–Väisälä frequency used in 1D+ model, i.e., the growth rate is overestimated in 1D+ model.
If we can keep the seed turbulent velocity in the pre-shocked region as same as that of 3D model, the compressed turbulent energy could become the level of the 3D simulation, that does not significantly alter the shock evolution.
Then the compressive model of 1D+ would provide more consistent result with 3D. Since there is no obvious way to keep the seed turbulent velocity small in the pre-shocked region, we prefer using the non-compressive case in order to mimic shock evolution of 3D.
We keep this issue in mind when developing the code in the future.

Based on the above argument, we consider the non-compressive case, which is unaffected by the compression, as the fiducial model that can mimic 3D shock evolution. In the following, we present the results of the non-compressive case by systematically changing the turbulence parameters.
In \cite{Couch2020}, $\dot{e}_{\rm HT}$ is described as shear production and they reported that the term is negligible.
The treatment would be practically similar to ours. Obviously, the current treatments are not perfect. Other sources of the turbulence and the amplification near the shock should be investigated in the future \citep[see][and references therein]{Abdikamalov2016}.
\subsection{Dependence of Mixing length parameter} \label{sec:mixing}

Here, we systematically vary the mixing length parameter in the non-compressive setup and 
present their dynamics. First, we show the dependence of the shock evolution on the mixing length parameter. Next, we show the velocity distribution of the turbulence and illustrate the effect of the mixing length parameter.

Since the source term of the turbulence is governed by the mixing length and proportional to $\alpha_\Lambda$ in Eq.~\eqref{eq:sourse_turb}, this parameter should play an essential role in the dynamics. $\alpha_\Lambda$ significantly affects the shock propagation.  In the models with larger $\alpha_\Lambda$, the stalled shock more easily revive. 
Here, we explain those issues one by one.
\begin{figure}
    \centering
    \includegraphics[width=.99\linewidth]{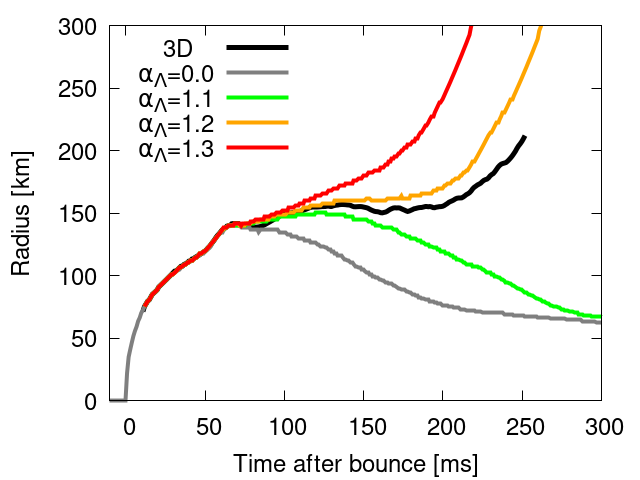}
    \caption{Shock evolution of $\alpha_\Lambda=0.0$, $1.1$--$1.3$.
    The color of the curve is green for 1.1, orange for 1.2, red for 1.3, and gray for 0.0. 
    The position of the average shock radius of 3D is plotted as black curve.}
    \label{fig:shock-mix}
\end{figure}
\begin{figure}
    \centering
    \includegraphics[width=.99\linewidth]{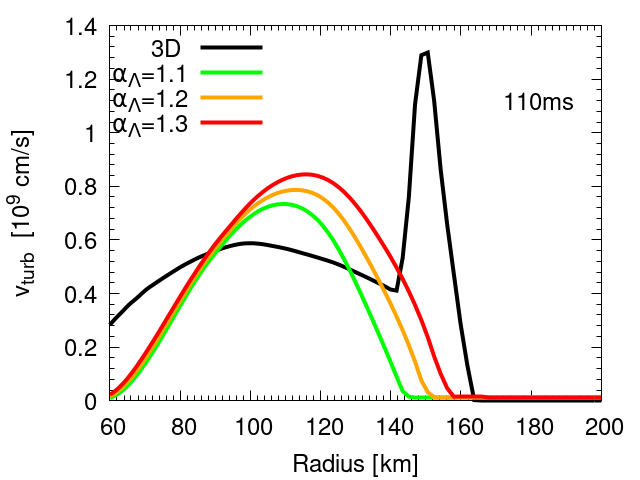}
    \caption{Radial profile of turbulent velocity for 120\,ms after core bounce for three models $\alpha_\Lambda = 1.1$--$1.3$. The color of the plot is as same as Figure~\ref{fig:shock-mix}. The black line is the turbulent velocity distribution in 3D.}
    \label{fig:vturb-mix}
\end{figure}

Figure~\ref{fig:shock-mix} shows the shock evolution of the models of $\alpha_\Lambda=0.0, 1.1$--$1.3$. The horizontal axis is the time from the core bounce, and the vertical axis is the shock radius.
The positions of the averaged shock radius of 3D are plotted by black line.
The model of $\alpha_\Lambda = 0.0 $ (gray line) is a model that does not employ multi-dimensional effects.
In the figure, the later time evolution is depicted, which is not shown in Figure~\ref{fig:CompShock}.

The model of $\alpha_\Lambda = 1.2$ can mimic the 3D model well. The shock revival becomes faster if $\alpha_\Lambda$ becomes larger.
In the model of $\alpha_\Lambda < 1.2$, the shock stagnates and fails to explode. On the other hand, for models $\alpha_\Lambda \geq 1.2$, the shock revives and propagates out of the iron core. The $\alpha_\Lambda$ monotonically enhances the explodability of the massive stars.  This tendency is roughly consistent with the previous study \citep[see Figure~2  of ][]{Couch2020}.

Next, we compare the distribution of turbulent velocity of \verb|1D+| with the 3D model.
Figure~\ref{fig:vturb-mix} shows the radial profile of turbulent velocity for 110\,ms after core bounce for three models $\alpha_\Lambda = 1.1$--$1.3$. The horizontal axis is radius. The color of the plot is green for 1.1, orange for 1.2, and red for 1.3 as same as Figure~\ref{fig:shock-mix}. The black line is the turbulent velocity distribution in 3D.

The turbulent velocity of \verb|1D+| peaks near gain radius, 100\,km (solid color curves).
The profile of the 3D model near the gain radius is similar to the \verb|1D+| models while the width of the turbulence is wider than \verb|1D+|.
As expected, the turbulent velocity increases monotonically as $\alpha_\Lambda$ increases. This trend is similar to \cite{Couch2020}. See their Figure~1. The corresponding peak turbulent pressure, $P_{\rm turb}$ in our model is $\sim 2 \times 10^{27}\,{\rm erg/cm^3}$ at 100-120\,km. That is larger than that in Figure~2 of \cite{Mabanta2019}, $\sim 0.8 \times 10^{27}\,{\rm erg/cm^3}$. Their assumed neutrino luminosity is $2.1\times 10^{52}\,{\rm erg/s}$ and is smaller than that of our model, $5\times 10^{52}\,{\rm erg/s}$. \cite{Murphy2013} argues the strength of turbulence is positively correlated with neutrino heating. The difference from their model is reasonable. Note that we use normal Reynolds decomposition while previous study may use different definitions. Our treatment in 3D may exaggerate the peak near the shock \citep[see footnote~2 of ][]{Murphy2011}.

\subsection{Diffusion parameters}\label{sec:diff}
The effective turbulence model has other parameters than $\alpha_\Lambda$.
The effect of the parameters has not been adequately investigated in the previous works (see Section~\ref{sec:introduction}).
We systematically change the diffusion constant $\alpha_u$ ($u= K$, $\epsilon$, $Y_e$).
The definition of the parameters is given in Eq.~\eqref{eq:diff}.
We fix $\alpha_\Lambda=1.2$
in the non-compressive setup as in Section~\ref{sec:comp}.

\begin{figure}
    \centering
    \includegraphics[width=.99\linewidth]{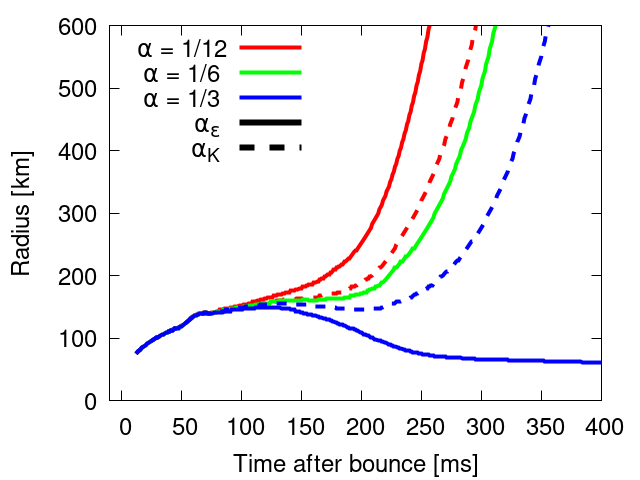}
    \includegraphics[width=.99\linewidth]{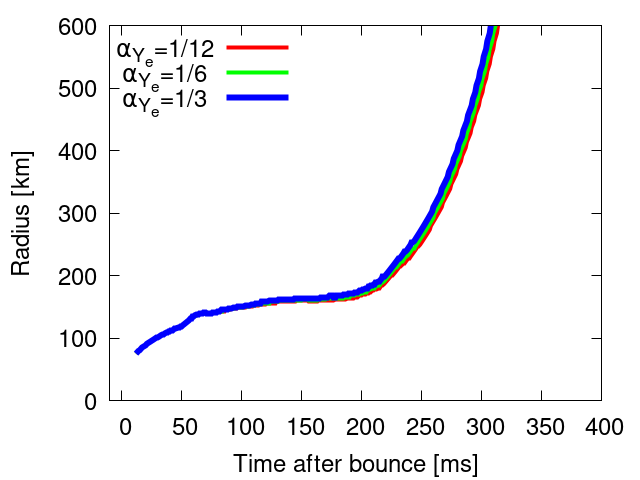}
    \caption{Shock evolution for different values of diffusion parameters. The horizontal axis is the time after core bounce and the vertical axis is radius. In the top panel, the green line is the fiducial model $(\alpha_\epsilon ,\alpha_K, \alpha_{Y_e})=(1/6 ,1/6,1/6)$. We change $\alpha_\epsilon$ (solid line) and $\alpha_K$ (dashed line) and the red line corresponds $\alpha_u=1/12$, and the blue line is $\alpha_u=1/3$. Green dashed line and solid line are overlapped.}
    The bottom panel is the same to the top panel but $\alpha_{Y_e}$ is changed. All lines are overlapped.
    \label{fig:diff_shock}
\end{figure}

Before going to the details, let us summarize the results in this section.
Interestingly, $\alpha_\epsilon$ and $\alpha_K$ also affect the hydrodynamic evolution. In particular, $\alpha_\epsilon$ has a significant impact on the explodability.
Figure~\ref{fig:diff_shock} compiles the shock evolution in different values of diffusion parameters. The horizontal axis is the time after core bounce and the vertical axis is radius.
The top panel shows the shock propagation for the fiducial model and the parameters $\alpha_\epsilon$ (solid line) and $\alpha_K$ (dashed line).  The green line is the fiducial model $(\alpha_\epsilon ,\alpha_K, \alpha_{Y_e})=(1/6 ,1/6,1/6)$, the red line is $1/12$, and the blue line is $1/3$.

As shown in the top panel of Figure~\ref{fig:diff_shock}, the shock evolution strongly depends on $\alpha_\epsilon$ and $\alpha_K$. Surprisingly, the model of $\alpha_\epsilon =1/3$ (blue line) fails to explode. 
The dynamics might be more sensitive to $\alpha_\epsilon$ than $\alpha_K$.
The shock revival time of $\alpha_\epsilon=1/12$ is earlier than $\alpha_K=1/12$. 
The time is delayed in $\alpha_K=1/3$ and $\alpha_\epsilon=1/3$ cannot show the shock revival.
On the other hand, $\alpha_{Y_e}$ does not change the shock dynamics (see the bottom panel).
Note that we switch off the turbulent effect in PNS by hand. In this setup, the diffusion of $\alpha_{Y_e}$ does not play any role.

\subsubsection{Diffusion of internal energy}
Here, we try to interpret why the smaller diffusion parameter, $\alpha_\epsilon$, make the shock revival easier. This is unexpected and has not been discussed in the literature.
 \cite{Yamasaki_2006} showed that the strong energy transport due to the turbulence reduces the critical neutrino luminosity that is necessary for the explosion, i.e., a larger diffusion parameter is favorable for the explosion. Why are our results contrary to this naive expectation?

\begin{figure}
     \centering
     \includegraphics[width=.9\linewidth,clip]{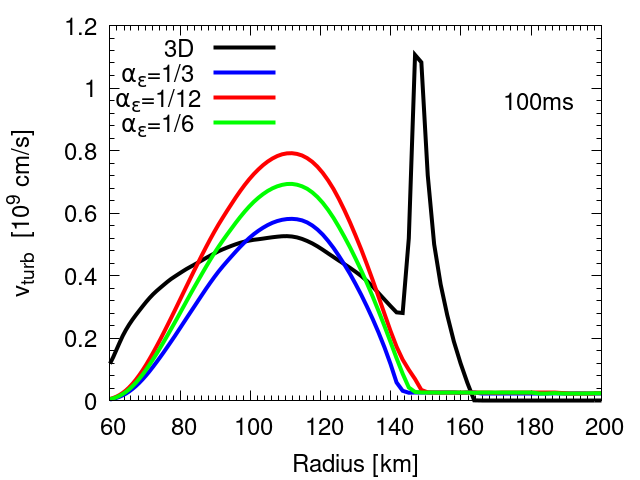}\\
     \includegraphics[width=.9\linewidth,clip]{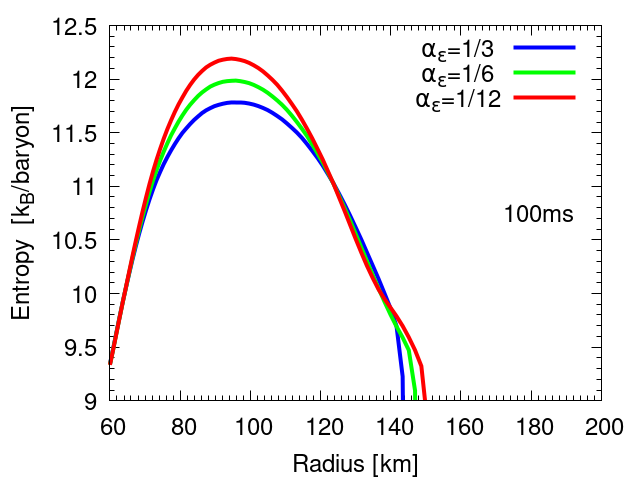}\\
     \includegraphics[width=.9\linewidth,clip]{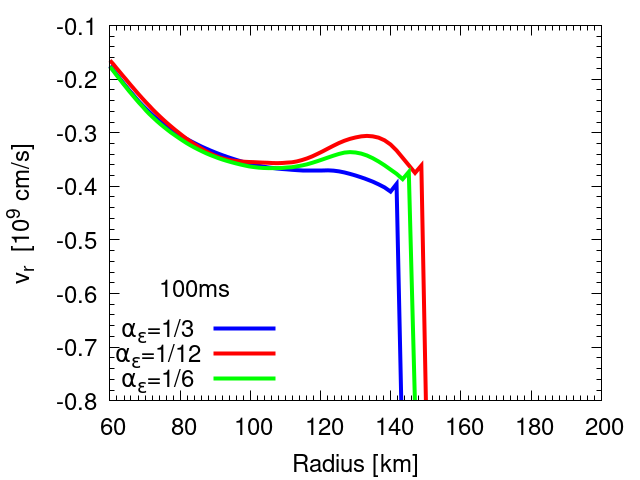}
     \caption{Radial profiles of turbulent velocity, entropy and radial velocity in different $\alpha_\epsilon$. From top to bottom panel, we show the radial profiles of turbulent velocity, entropy and radial velocity at 100\,ms core bounce.     
     We change the parameter $\alpha_\epsilon = 1/3$, $1/6$, and $1/12$ and 
     the colors of the curves are blue for 1/3, green for 1/6, and red for 1/12 as same as Figure~\ref{fig:diff_shock}.}
     \label{fig:vturb-e}
\end{figure}
The earlier shock revival is correlated with longer advection time, $\tau_{\rm adv}$ \citep[e.g.,][]{Buras2006b}.
In the following paragraphs, we elucidate the underlying reasons why the advection time is longer when the diffusion constant becomes smaller.
Figure~\ref{fig:vturb-e} shows the radial profile of turbulent velocity, entropy and velocity for 100\,ms after core bounce for three models $\alpha_\epsilon = 1/3$, $1/6$, and $1/12$. The horizontal axis is radius.  The color of the curve is blue for 1/3, green for 1/6, and red for 1/12 as same as Figure~\ref{fig:diff_shock}. 

Examining the top panel and the middle panel, it becomes evident that reducing the diffusion term leads to an increase in turbulent velocity within the range of 60\,km to 140\,km. This is because turbulent energy generation depends on the  Brunt–Väisälä  frequency, which is associated with the entropy gradient. Since the diffusion weakens the entropy gradient, a larger turbulent energy generation is expected in the lower diffusion constant.

Consequently, the enhanced turbulent velocity, or stronger turbulent pressure, contributes to reducing the post-shock radial velocity.  As observed in the bottom panel, specifically around the region of 120\,km to 150\,km, the absolute value of the radial velocity is smaller for the red curve (corresponding to $\alpha_\epsilon = 1/12$) compared to the green and blue curves (representing $\alpha_\epsilon = 1/6$ and $1/3$, respectively). Although the difference is subtle in this particular snapshot, it becomes more pronounced over time.

\begin{figure}
     \centering
     \includegraphics[width=.9\linewidth,clip]{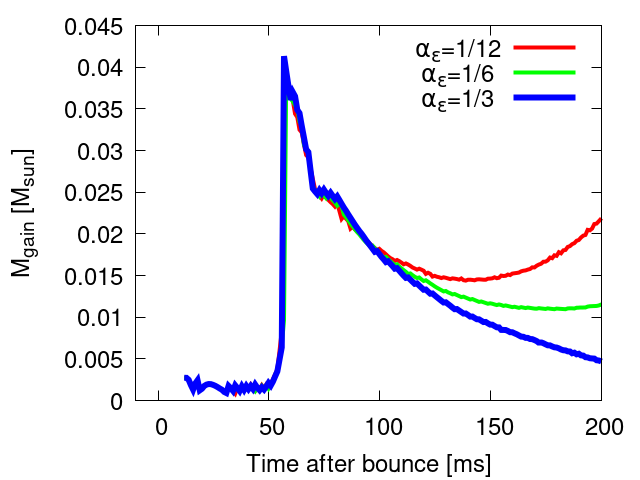}\\
     \includegraphics[width=.9\linewidth,clip]{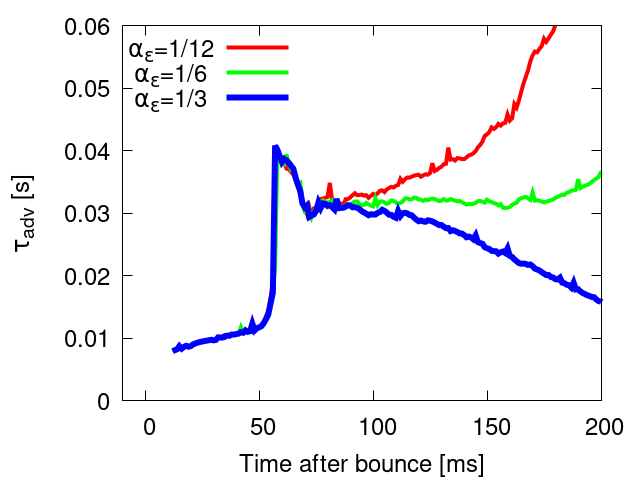}
     \caption{Time evolution of gain mas and advection timescale for $\alpha_\epsilon = 1/3$, $1/6$, and $1/12$. The top and bottom panel shows the time evolution of gain mass $[M_\odot]$ and $\tau_{\rm adv}$, respectively. 
     Horizontal axis is time after core bounce in the top and bottom panel.
     The colors of the curves are blue for 1/3, green for 1/6, and red for 1/12 as same as Figure~\ref{fig:diff_shock}.}
     \label{fig:time-e}
\end{figure}
As a consequence of that, the gain mass becomes larger in the less diffusive models. Figure \ref{fig:time-e} show the evolution of gain mass $[M_\odot]$ (top) and advection timescale (bottom), $\tau_{\rm adv}=M_{\rm gain}/\dot{M}_{\rm acc}$. The horizontal axis is time after core bounce. The color of the curve is blue for 1/3, green for 1/6, and red for 1/12 as same as Figure~\ref{fig:diff_shock}.
 
In the top panel, the difference in the gain mass is apparent from 100\,ms. The detailed radial profile at that time is shown in Figure~\ref{fig:vturb-e}. The model with $\alpha_\epsilon=1/12$, which has a slow radial velocity profile, has a larger gain mass. On the other hand, the $\alpha_\epsilon=1/3$ model has a smaller gain mass. The larger gain mass corresponds to longer advection timescale as shown in the bottom panel.

To summarize, the lower diffusion constant of internal energy keeps the entropy gradients, leading to 
efficient turbulence generation. The high turbulent pressure decelerates the radial velocity in the post shock region and consequently the gain mass becomes larger and advection timescale longer. That results in the earlier shock revival.

\subsubsection{Diffusion of turbulent energy}

\begin{figure}
     \centering
     \includegraphics[width=.9\linewidth,clip]{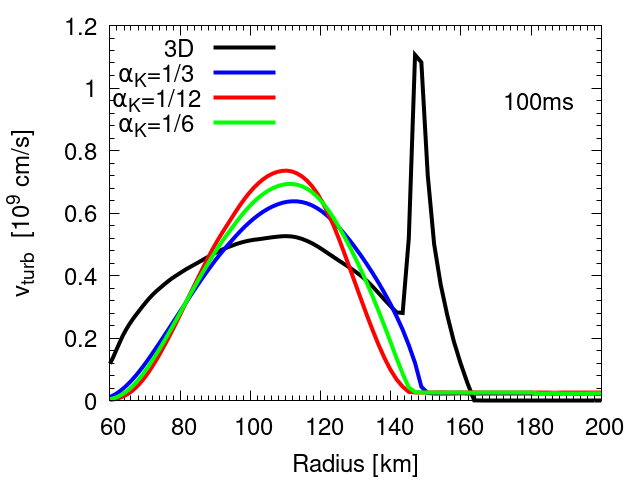}\\
     \includegraphics[width=.9\linewidth,clip]{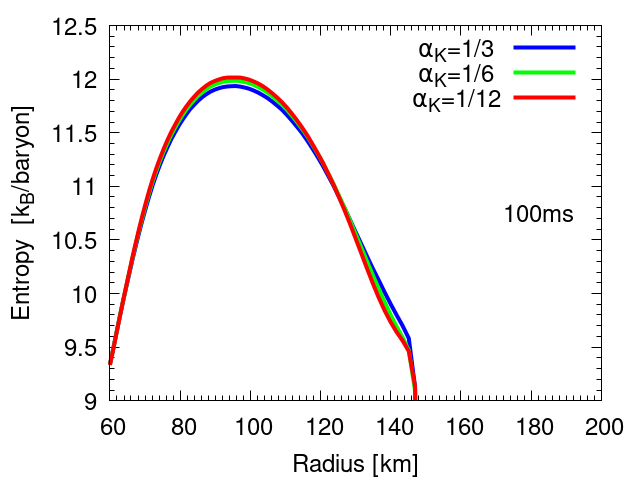}\\
     \includegraphics[width=.9\linewidth,clip]{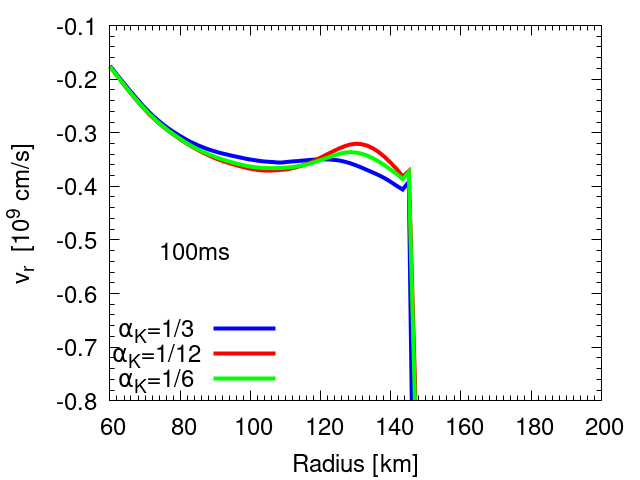}
     \caption{Radial profiles of turbulent velocity, entropy and radial velocity in different $\alpha_K$.
      Same as Figure~\ref{fig:vturb-e} but the dependence on $\alpha_K$ is shown.
     The turbulent velocity, entropy and radial velocity at 100\,ms core bounce are shown in the top, middle and bottom panels, respectively. The colors of the curves are blue for $\alpha_K = 1/3$, green for 1/6, and red for 1/12 as same as Figure~\ref{fig:diff_shock}.}
     \label{fig:vturb-k}
\end{figure}

We examine the impact of turbulent energy diffusion, $\alpha_K$, in a similar vein to the previous section, which discusses the internal energy diffusion. Figure~\ref{fig:vturb-k} shows the radial profile of turbulent velocity (top), entropy (middle), and velocity (bottom) for 100\,ms after core bounce for three models $\alpha_K = 1/3$, $1/6$, and $1/12$
(see Figure~\ref{fig:vturb-e} for the case of $\alpha_\epsilon$). The horizontal axis is radius. The color of the curve is blue for 1/3, green for 1/6, and red for 1/12 as same as Figure~\ref{fig:diff_shock}. 

From the middle panel, it is evident that reducing the diffusion constant, $\alpha_K$, does not alter the entropy profile. 
Moving to the top panel, the amplitude of turbulent velocity is higher if the diffusion constant is smaller
and the width becomes narrower in the range of 80\,km to 150\,km.
The diffusion blunts the turbulent energy distribution.
While the diffusion of internal energy, $\alpha_\epsilon$, influences the turbulence generation by changing the entropy distribution as explained by Figure~\ref{fig:vturb-e}, the diffusion of turbulent energy does not affect the generation of turbulence, but just alters the turbulent velocity distribution.

The profile of the radial velocity is affected by the turbulent pressure. In the bottom panel, the less diffusive model (red curve) shows the more decelerated profile. The model has the steepest gradients in the middle panel and leads to the highest turbulent pressure gradient force which efficiently slow down the radial velocity.

\begin{figure}
     \centering
     \includegraphics[width=.9\linewidth,clip]{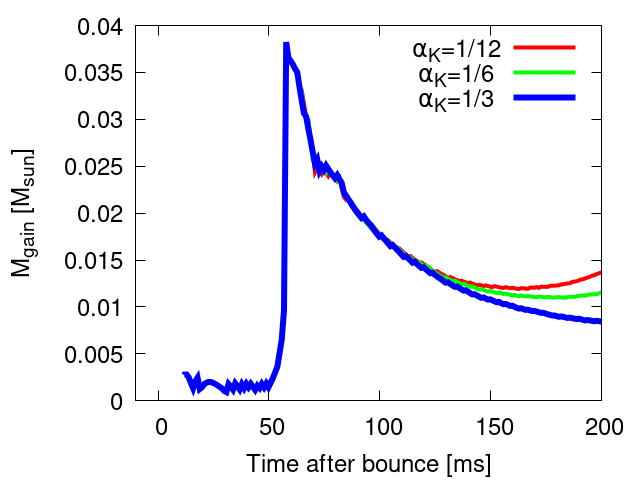}\\
     \includegraphics[width=.9\linewidth,clip]{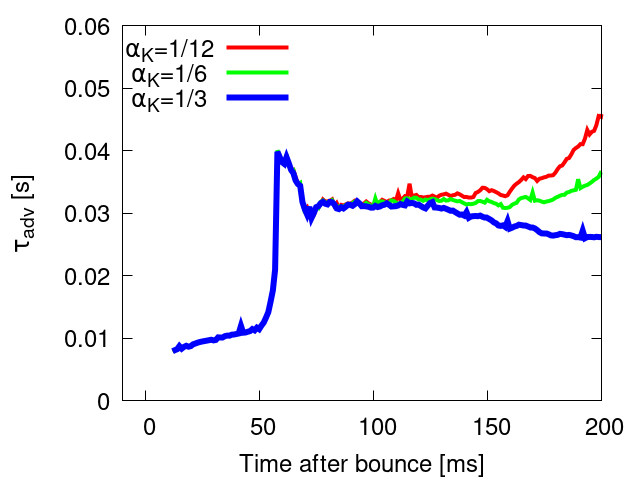}
     \caption{Time evolution of gain mas and advection timescale for $\alpha_K = 1/3$, $1/6$, and $1/12$.
     Same as Figure~\ref{fig:time-e} but the dependence on $\alpha_K$ is shown.
     he top and bottom panel shows the time evolution of gain mass $[M_\odot]$ and $\tau_{\rm adv}$, respectively. 
     The colors of the curves are blue for $\alpha_K = 1/3$, green for 1/6, and red for 1/12 as same as Figure~\ref{fig:diff_shock}.}
     \label{fig:time-k}
\end{figure}

Consequently, the gain mass becomes larger in the less diffusive model as same as effects of diffusion of internal energy, $\alpha_\epsilon$. Figure~\ref{fig:time-k} shows Evolution of gain mass $[M_\odot]$ (top) and advection timescale (bottom).  The horizontal axis is time after core bounce. The color of the curve is blue for 1/3, green for 1/6, and red for 1/12 as same as Figure~\ref{fig:diff_shock}. In the bottom panel of Figure~\ref{fig:vturb-k}, the model with $\alpha_K=1/12$, which has a slow radial velocity profile, has a larger gain mass and longer advection timescale. On the other hand, the $\alpha_K=1/3$ model has a smaller gain mass and shorter advection timescale. 
Compared to the bottom panel of Figure~\ref{fig:vturb-e}, the variation in gain mass is small.

The diffusion of turbulent energy profoundly impacts gain mass, advection timescale, and the evolution of the shock.
The stronger diffusion makes the radial profile of the turbulent velocity flatter leading to smaller turbulent pressure gradient force.  
The radial velocity of the post shock region is less affected and maintained at higher values. As a result, the gain mass becomes smaller and the advection timescale shorter.

\section{Summary and Discussion}
We have performed 1D simulations with phenomenological turbulent effects (so-called \verb|1D+| simulation).
The impact of several turbulent terms are investigated with a $12 M_\odot$ progenitor model. First the effect of the compression or shear production is focused and then the effects of the mixing length parameter $\alpha_\Lambda$ and the diffusion parameters $\alpha_\epsilon$, $\alpha_K$, $\alpha_{Y_e}$ are investigated.

We have compared the evolution of the shock among the models with the compression term, $\Dot{e}_{\rm HT}$ and without it.
The model without it agree with the 3D simulations. 
The enhancement of turbulent energy due to the compression, which is defined by Eq.~\eqref{eq:C-gain}, is too large at early phase of shock propagation in our \verb|1D+| simulations. 

We have adjusted the mixing length parameter so that can mimic the evolution of a 3D shock in the non-compression case. Our model of $\alpha_\Lambda =1.2$ mimics well the shock evolution. The turbulent velocity profile also shows good agreement with the 3D turbulent velocity profile, especially near the gain radius where neutrino heating is strong.

We analyze the effect of diffusion parameters $\alpha_\epsilon,\alpha_K, \alpha_{Y_e}$. The standard model is the non-compressive case with $\alpha_\Lambda=1.2$.
Among them, the diffusion of internal energy, $\alpha_\epsilon$, most significantly affect the evolution of shock.
The diffusion coefficients of internal energy,$\alpha_\epsilon$, and  turbulent energy, $\alpha_K$, also affect the explodability. The smaller diffusion makes the shock revival faster. Our comparison between the two reveals that the diffusion coefficients of internal energy has a greater impact.(see Figure~\ref{fig:diff_shock}).
$\alpha_K$ has relatively smaller impact and $\alpha_{Y_e}$ has no impact on the explosion in our setup. 

We analyze the reason why the lower diffusion of $\alpha_\epsilon$ makes the explosion stronger.
It is contrary to the expectations of \cite{Yamasaki_2006}, which demonstrate how the energy diffusion helps the shock revival.
The lower diffusion of $\alpha_\epsilon$, the greater the turbulent energy is generated.
The turbulent pressure prevents the accretion and  the advection time becomes longer.
Since $\alpha_K$  does not affect turbulent generation, the impact becomes smaller.

Afterward, we address caveats in this study.
Since \verb!1D+! approach is a relatively new idea, the method has not been maturely established yet.
The method based on Reynolds decomposition should include the turbulent mass flux \citep{Mueller2019}, which is ignored in this study \citep[it is same as][]{Mabanta2019,Couch2020,Boccioli2021}. On the other hand, the method based on Favre decomposition causes a different problem (see Appendix~\ref{sec:formalism}).
More sophisticated methods should be developed.In any case, a detailed comparison would be necessary to understand the difference.

While this study focus on only the convection in the gain region, 
the convection in proto-neutron star (PNS) would enhance the neutrino luminosity and hence affect the dynamics of the shock
\citep{Keil1996,Buras2006b,Nagakura2020_PNS_CONVECTION}.
Phenomenological treatments on this should be investigated \citep[e.g.,][]{Hudepohl2014}.
Though $\alpha_{Y_e}$ does not change the dynamics in our study since we switch off the turbulent effect in PNS by hand, this parameter would be important in the study of PNS convection.

Though we have investigated the effect of the shock using two extreme cases, compressive and non-compressive.
The treatment is not perfect.
If the convection at the early phase is significant, that may change the position of the stalled shock \citep{Nagakura2018,Harada2020}.
The strength of the convection may depend on the detail of the initial perturbation seed \citep{Kazeroni2018,Kazeroni2020}. We need more careful argument on this issue \citep{Takahashi2014,Takahashi2016,Huete2018,Huete2020}.

One essential question is "How accurately does the phenomenological model reproduce the dynamics of the multi-dimensional simulation?".
First we need to evaluate the parameters in multi-dimensional simulations.
One difficulty with this issue is the numerical resolution of the multi-dimensional simulations.
As \cite{Nagakura2019resolution} reported, the numerical resolution significantly changes the turbulent properties.
We need high-resolution simulation to evaluate the parameters \citep[e.g.,][]{Radice2016,Melson2020}.

Perhaps we have to take a significantly different approach to treat convective turbulence.
As discussed in \cite{Murphy2013}, the turbulent convection in the gain region is driven by the neutrino heating, not directly by negative entropy gradients as in Eq.~\eqref{eq:sourse_turb}, which is valid only in the linear phase.
To precisely model the production term, we may have to employ global modeling \citep[e.g.,][]{Mabanta2019} instead of the local modeling, such as MLT \citep[see also][for taking non-local effect in the local modeling]{yokoi2022}.
In the global modeling, the turbulent effect is not given by the local variables and needs a global structure of the flow,
which is characterized by spatial integration of the variables, analytical closure, and empirical relations \citep{Murphy2011,Mabanta2019}.
Though we have taken local modeling in this study just due to its simpleness, we need further effort to go to more detailed modeling in the future. 

The next important point is to examine whether the flux due to turbulence is consistant with 3D. As long as the gradient-diffusion approximation used in this study is correct, our results should be justified, but more detailed studies are needed. In addition to flux, turbulent pressure is another physical quantity that is important when evaluating turbulence.

\verb!1D+! simulations are robust tools for parametric studies of the core-collapse supernovae, though there are still many problems.
A variety of applications are possible, for example;
progenitor dependence \citep{Couch2020};
neutrino, gravitational wave, optical light curves \citep{Warren2020,Barker2022};
detailed neutrino emission \citep{Suwa2019,Mori2021};
EOS dependence \citep{Yasin2020,Nakazato2021};
BH formation \citep{Schneider2020};
quark star phase transition \citep{Fischer2020};
emission of axion-like particle \citep{Fischer2016} and so on.
We hope that this study will take us one step closer to understanding a vast and untapped area of the research field.

\section*{Acknowledgements}

We thank Y. Masada, N. Yokoi and B. M\"{u}ller for stimulating discussions.
This work was partly supported by JSPS KAKENHI
Grant Number (JP17H06364, 
JP18H01212, 
JP21H01088, 
JP22H01223  
and JP23K03400). 
This research was also supported by MEXT as “Program for Promoting 
researches on the Supercomputer Fugaku” (Structure and Evolution of the Universe Unraveled by Fusion of Simulation and AI; Grant Number JPMXP1020230406) and JICFuS.
The National Institutes of Natural Sciences 
(NINS) program for cross-disciplinary
study (Grant Numbers 01321802 and 01311904) on Turbulence, Transport,
and Heating Dynamics in Laboratory and Solar/Astrophysical Plasmas:
``SoLaBo-X" supports this work.
Numerical computations were carried out on PC cluster at Center for Computational Astrophysics, National Astronomical Observatory of Japan.

\section*{Data Availability}
The data underlying this article will be shared on reasonable request to the corresponding author.



\bibliographystyle{mnras}
\bibliography{references,refinarXiv} 




\appendix

\section{Comparison of the energy equation between Reynolds and Favre decomposition}\label{sec:formalism}

\begin{figure*}
 \centering
 \includegraphics[width=0.45\linewidth,clip]{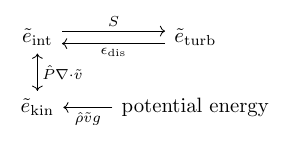}
 \includegraphics[width=0.45\linewidth,clip]{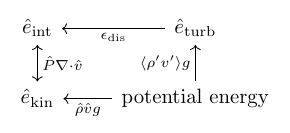}
 \caption{Schematic diagram for global energy balance in turbulent convection. The diagrams of the energy flow for Favre and Reynolds decomposition are depicted in Left and Right panels, respectively. }
 \label{fig:energy-flow}
\end{figure*}

We explain the pros and cons of our formalism shown in Section~\ref{sec:numericalsetup}.
Two kinds of approaches are proposed in the literature to construct phenomenological turbulence models : one is based on Reynolds decomposition \cite[e.g.][]{Couch2020}, and the other is based on Favre decomposition \citep{Mueller2019}. 
In this context, Reynolds decomposition treats the deviation from Reynolds averaged quantity, $\hat{A}=\langle A \rangle = \iint A(r,\theta,\phi) \sin\theta{\rm d}\theta{\rm d}\phi/4\pi$, where the left-hand side is a function of $r$.
Favre decomposition considers the deviation from Favre average, $\tilde{A} = \langle \rho A \rangle /\langle \rho \rangle $, i.e., the mass-weighted average is performed.

One of the essential differences appears in the mass conservation equation.
\begin{linenomath*}
\begin{align}
    \partial_t \langle \rho \rangle & + \nabla_r \left(\hat{\rho}\hat{v}+\langle \rho^\prime v^\prime \rangle\right) =0, \label{eq:rynld_mfx}\\ 
    \partial_t \langle \rho \rangle & + \nabla_r \left(\tilde{\rho}\tilde{v}\right) =0, \label{eq:favre_mfx}
\end{align}
\end{linenomath*}
where prime means the deviation from Reynolds averaged quantity.
In Reynolds decomposition, the total mass flux is written as the sum of the mean flux, $\hat{\rho}\hat{v}$ and the turbulent flux $\langle \rho^\prime v^\prime\rangle$.
In Favre decomposition, the total mass flux is the same as the mean flux, $\tilde{\rho}\tilde{v}$.
The turbulent mass flux does not appear.
The treatment of the mass flux causes a significant difference in the source term of energy transport.

\cite{Mueller2019} takes advantage of the simple description of Favre decomposition to construct energy conserving formalism:
\begin{linenomath*}
\begin{align}
    \partial_t \tilde{e}_{\rm tot} +\cdots =&  \hat{\rho} \tilde{v} g \label{eq:favre_etot}\\
    \partial_t \tilde{e}_{\rm kin} +\cdots =& \hat{\rho} \tilde{v} g -\tilde{v}\cdot\nabla \left(\hat{P}+P_{\rm turb}\right)\label{eq:favre_ekin}\\
    \partial_t \tilde{e}_{\rm turb} +\cdots =& S -\epsilon_{\rm dis} -P_{\rm turb}\nabla\cdot \tilde{v}\label{eq:favre_etub}\\
    \partial_t \tilde{e}_{\rm int} +\cdots =& -S +\epsilon_{\rm dis} -\hat{P}\nabla\cdot \tilde{v} \label{eq:favre_eint}
\end{align}
\end{linenomath*}
Clearly, the sum of Eqs.~\eqref{eq:favre_ekin}--\eqref{eq:favre_eint} is identical to Eq.~\eqref{eq:favre_etot}.

We point out a problem of \cite{Mueller2019}'s formalism.
$S$ in the Eq.~\eqref{eq:favre_etub} generates the turbulent energy and is written as 
$S=\langle \rho^\prime v^{\prime\prime} \rangle g$, which is buoyant driving.
Here is where it gets weird, i.e., the effect of the buoyancy appears in the equation of the internal energy, Eq.~\eqref{eq:favre_eint}. 
Usually, the buoyancy changes only the kinetic energy and does not change the internal energy.
Though \cite{Mueller2019} is conscious of this problem and tries to justify their formalism in section 3.3 assuming pressure gradient as gravitational force,
we have a different, simpler view based on \cite{Viallet2013} (see their Figure 20).
The energy transport between the turbulent energy and internal energy should be pressure fluctuations or pressure-dilatation, $W_p\sim \langle P^\prime \nabla\cdot v^{\prime\prime} \rangle $.
The formalism using Favre decomposition would be suitable when the source term of the turbulent energy is pressure fluctuation \citep[see also][]{Scannapieco2008}.
We visualize the energy balance in Favre decomposition in Figure~\ref{fig:energy-flow} left panel.

Similar equations can be written in Reynolds decomposition:
\begin{linenomath*}
\begin{align}
    \partial_t \hat{e}_{\rm tot} +\cdots  =& \left( \hat{\rho} \hat{v} + \langle \rho^{\prime} v^{\prime} \rangle \right) g \label{eq:rynld_etot}\\
    \partial_t \hat{e}_{\rm kin} +\cdots =& \hat{\rho} \hat{v} g -\hat{v} \cdot \nabla \left(\hat{P}+P_{\rm turb}\right) \label{eq:rynld_ekin} \\
    \partial_t \hat{e}_{\rm turb} +\cdots =&  \langle \rho^{\prime} v^{\prime} \rangle g -\epsilon_{\rm dis}  -P_{\rm turb}\nabla\cdot \hat{v} \label{eq:rynld_etub}\\
    \partial_t \hat{e}_{\rm int} +\cdots =& \epsilon_{\rm dis}   -\hat{P}\nabla\cdot \hat{v}\label{eq:rynld_eint}
\end{align}
\end{linenomath*}
Again, the sum of Eqs.~\eqref{eq:rynld_ekin}--\eqref{eq:rynld_eint} is identical to Eq.~\eqref{eq:rynld_etot}.
In Reynolds decomposition, the turbulent energy is naturally generated by buoyant force,  $\langle \rho^{\prime} v^{\prime} \rangle g$.
This point is a virtue of Reynolds decomposition.
We visualize the energy balance in Reynolds decomposition in Figure~\ref{fig:energy-flow} right panel.

The formalism with Reynolds decomposition has weak points.
First, the definition of turbulent energy is complex, i.e.,
$\hat{e}_{\rm turb}=\langle \rho^\prime {v^\prime}^2 \rangle/2+\hat{\rho}\langle{v^\prime}^2\rangle/2 +\langle \rho^\prime v^\prime \rangle  \hat{v} $.
Second, we should additionally consider turbulent mass flux in Eq.~\eqref{eq:rynld_mfx},
which is ignored in our formalism.
\cite{Mueller2019} criticizes this point.
Since the potential energy changes by Eq.~\eqref{eq:rynld_mfx}, it is important to include the turbulent mass flux in the equation.

We summarize the pros and cons of the two formalism.
{\it Pros}: Favre decomposition makes 
the equations simpler, i.e., the definition of turbulent energy is straight forward, and 
the equation of mass conservation is easily treated.
{\it Cons}: turbulent energy generation term in Favre decomposition is limited to that of pressure fluctuation.
The employment of buoyancy force causes strange energy transfer between turbulent and internal energies.
{\it Pros}: Reynolds decomposition distinguishes the mean mass flux and turbulent mass flux (see Eq.~\eqref{eq:rynld_mfx}).
Thus the energy generation by turbulent mass flux naturally appears in the turbulent energy equation, Eq.~\eqref{eq:rynld_etub}.
{\it Cons}: The definition of turbulent energy becomes complex, and good treatment of convective mass flux 
in Eq.~\eqref{eq:rynld_mfx} has not been known. 


\bsp	
\label{lastpage}
\end{document}